\documentclass[12pt,preprint]{aastex}

\shortauthors{Bower et al.}
\shorttitle{Millimeter Polarimetry of LLAGN}
\begin{document}

\newcommand\degd{\ifmmode^{\circ}\!\!\!.\,\else$^{\circ}\!\!\!.\,$\fi}
\newcommand{\etal}{{\it et al.\ }}
\newcommand{\uv}{(u,v)}
\newcommand{\rdm}{{\rm\ rad\ m^{-2}}}
\newcommand{\msuny}{{\rm\ M_{\sun}\ y^{-1}}}
\newcommand{\mylesssim}{\stackrel{\scriptstyle <}{\scriptstyle \sim}}
\newcommand{\lsim}{\stackrel{\scriptstyle <}{\scriptstyle \sim}}
\newcommand{\gsim}{\stackrel{\scriptstyle >}{\scriptstyle \sim}}
\newcommand{\sci}{Science}
\newcommand{\sgr}{PSR J1745-2900}
\newcommand{\sgra}{Sgr~A*}
\newcommand{\kms}{\ensuremath{{\rm km\,s}^{-1}}}
\newcommand{\masy}{\ensuremath{{\rm mas\,yr}^{-1}}}

\def\kbar{{\mathchar'26\mkern-9mu k}}
\def\totd{{\mathrm{d}}}


\title{What is the Hidden Depolarization Mechanism in Low Luminosity AGN?}
\author{
Geoffrey C.\ Bower,\altaffilmark{1}
Jason Dexter,\altaffilmark{2}
Sera Markoff,\altaffilmark{3}
Ramprasad Rao,\altaffilmark{1}
R.~L.~Plambeck\altaffilmark{4} 
}

\altaffiltext{1}{Academia Sinica Institute of Astronomy and Astrophysics, 645 N. A'ohoku Place, Hilo, HI 96720, USA; gbower@asiaa.sinica.edu.tw}
\altaffiltext{2}{Max Planck Institute for Extraterrestrial Physics, Giessenbachstr. 1, 85748 Garching, Germany}
\altaffiltext{3}{Anton Pannekoek Institute for Astronomy, University
  of Amsterdam, Science Park 904, 1098 XH Amsterdam, The Netherlands}
\altaffiltext{4}{Radio Astronomy Laboratory, University of California, Berkeley, CA 94720-3411, USA}

\begin{abstract}
Millimeter wavelength polarimetry of accreting black hole systems can provide a tomographic probe 
of the accretion flow on a wide range of linear scales.  We searched for linear polarization in
two low luminosity active galactic nuclei (LLAGN), M81 and M84, using the Combined Array for Millimeter
Astronomy (CARMA) and the Submillimeter Array (SMA).  We find upper limits of $\sim 1 - 2\%$ averaging
over the full bandwidth and with a rotation measure (RM) synthesis technique.  These low polarization
fractions, along with similar low values for LLAGN M87 and 3C84, suggest that LLAGN have qualitatively
different polarization properties than radio-loud sources and Sgr A*.  If the sources are intrinsically polarized
and then depolarized by Faraday rotation then we place lower limits on the RM of a few times $10^7 \rdm$ for the full
bandwidth case and $\sim 10^9 \rdm$ for the RM synthesis analysis.  These limits are inconsistent with or marginally
consistent with expected accretion flow properties.  Alternatively, the sources may be 
depolarized by cold electrons within a few Schwarzschild radii from the black hole, as suggested by numerical models.
\end{abstract}


\section{Introduction}

Models of black hole accretion span many orders of magnitude in linear scale but
observational constraints tend to probe only very narrow ranges, predominantly 
in regions where emission arises \citep{2014ARA&A..52..529Y}.  Faraday rotation provides a unique line-of-sight
probe that can measure the integral properties of accretion flows over a wide
range of radii and physical conditions.  A magnetized plasma rotates the plane of linear polarization by an angle that
is proportional to the rotation measure (RM), which is the 
line-of-sight integral of the electron density and the
parallel component of the magnetic field.  In cases where linearly polarized emission arises
near the black hole, Faraday rotation can probe scales ranging from a few to thousands
(or more) of Schwarzschild radii ($R_S$).

Millimeter wavelength observations are well-suited to probing accretion flows
around supermassive black holes.  At these
wavelengths, emission regions tend to be compact and originate from near the
black hole.  Further, the short wavelengths provide sensitivity to large RMs $\gsim 10^6 \rdm$,
which are characteristic of accretion flow models for many systems.  At longer wavelengths,
intrinsically polarized sources will be quickly depolarized through bandwidth or beam
depolarization mechanisms \citep[e.g.,][]{1999ApJ...521..582B}.

Millimeter wavelength polarimetric probes of the accretion flow were first developed for Sgr A* \citep{2003ApJ...588..331B,2007ApJ...654L..57M} and
have since been applied to M87 \citep{2014ApJ...783L..33K} and 3C 84 \citep{2014ApJ...797...66P}.  In the case
of Sgr A*, the detected RM $\sim -5 \times 10^5 \rdm$ sets a constraint on the accretion
rate of $\dot{M} \sim 10^{-8 \pm 1} \msuny$ and rules out canonical, loss-free
advection dominated accretion flows \citep[ADAFs;][]{1995ApJ...452..710N}.  Time
variability of the RM can be used as a tomographic probe of structures in the accretion flow
\citep{2005ApJ...618L..29B,2011MNRAS.415.1228P}.

In this paper, we apply the technique of millimeter wavelength polarimetry to two
nearby low luminosity AGN (LLAGN), M81 and M84.  Both of these sources
have been studied extensively across the electromagnetic spectrum 
\citep[e.g.,][]{2000ApJ...532..895B,2004AJ....127..119L,2008ApJ...681..905M,2015ApJ...811L...6B,2013MNRAS.432..530R}.
Dynamical black hole masses have been determined for both M81 and M84 to be
$6.5 \times 10^7 M_\odot$ and
$8.3 \times 10^8 M_\odot$, respectively 
\citep{2013ARAA..51..511K}.
In Section 2, we present observations obtained with the Combined Array for Research in Millimeter-wave Astronomy (CARMA) and the Submillimeter Array (SMA).
In Section 3, we present our results and analysis.  In Sections 4 and 5, we discuss
these results and provide our conclusions.

\section{Observations}

\subsection{CARMA Observations of M81}

M81 data were obtained with the CARMA D-array on 2013 Feb 18 in good weather
(3mm precipitable water).  Observations of M81 were interleaved with those of
the phase calibrator, 0958+655; 3C 84 was used as a passband calibrator.  
The synthesized beam size on M81 was $2.6 \times 2.4$ arcsec in position
angle 58 deg East of North.  The
8~hour observing track covered a parallactic angle range of 160 degrees (-100
to -180/+180 to +100) for M81.  The dual-polarization receivers are equipped
with broadband waveguide circular polarizers and orthomode transducers,
allowing simultaneous observations of right (R) and left (L) circular
polarization \citep{2015JAI.....450005H}.  The CARMA correlator was configured
to process two 2~GHz wide frequency bands, centered at 216.6~GHz in the
receivers' lower sideband and 231.1~GHz in the upper sideband.  The correlator
provided 0.0104~GHz frequency resolution for all four cross-polarizations
(RR,LL,RL,LR) 

Polarization leakage corrections were derived from a 5-hour observation of
3C 279 one month later, on 2013 Mar 18, that covered a 75 degree range in
parallactic angle.  The observing frequency and correlator configuration were
identical to those used for M81.  There are no moving parts in the dual
polarization receivers, and the leakage corrections are known to be stable over
periods of many months.  The leakage corrections compensate for cross-coupling
between the R and L channels.  As discussed by \citet{2015JAI.....450005H},
reflections within the receivers cause ripples on scales of $\sim 100$~MHz on
some of the telescopes.  For normal broadband observations of dust polarization
\citep[e.g.,][]{2014ApJS..213...13H} these ripples average out, but for RM
synthesis we take care to derive and apply the leakage corrections
independently for each 0.052~GHz wide frequency interval before calculating
Stokes I,Q,U,V for that interval.

\subsection{SMA Observations of M84}

The SMA data on M84 were taken on three different epochs, 30 Jan 2016, 21 Feb 2016, and 29 Feb 2016. The weather conditions were excellent on all three days with the 225 GHz atmospheric opacity between 0.02 and 0.04 (precipitable water vapor below 0.5 mm to 1 mm). The primary phase calibrator was 3C 273 which was also used to calibrate the passband and instrumental polarization. The observing time on M84 on each of the days was $\sim$7 hours while the parallactic angle coverage was $\sim$160 degrees (-90 degrees to +70 degrees). The SMA observed in a double sideband mode with the upper sideband covering a frequency range from 230.9 GHz to 234.9 GHz while the lower sideband covered a frequency range from 218.9 GHz to 222.9 GHz. The correlator was set to a frequency resolution of $\sim$3 MHz. The instrumental polarizations or leakages were derived from 3C 273 which was observed simultaneously with the target observations of M84 \citep{2008SPIE.7020E..2BM}. The parallactic angle coverage was  similar to that of the target M84 (i.e., 160 degrees). 

\begin{figure}[p!]
\includegraphics[width=0.5\textwidth]{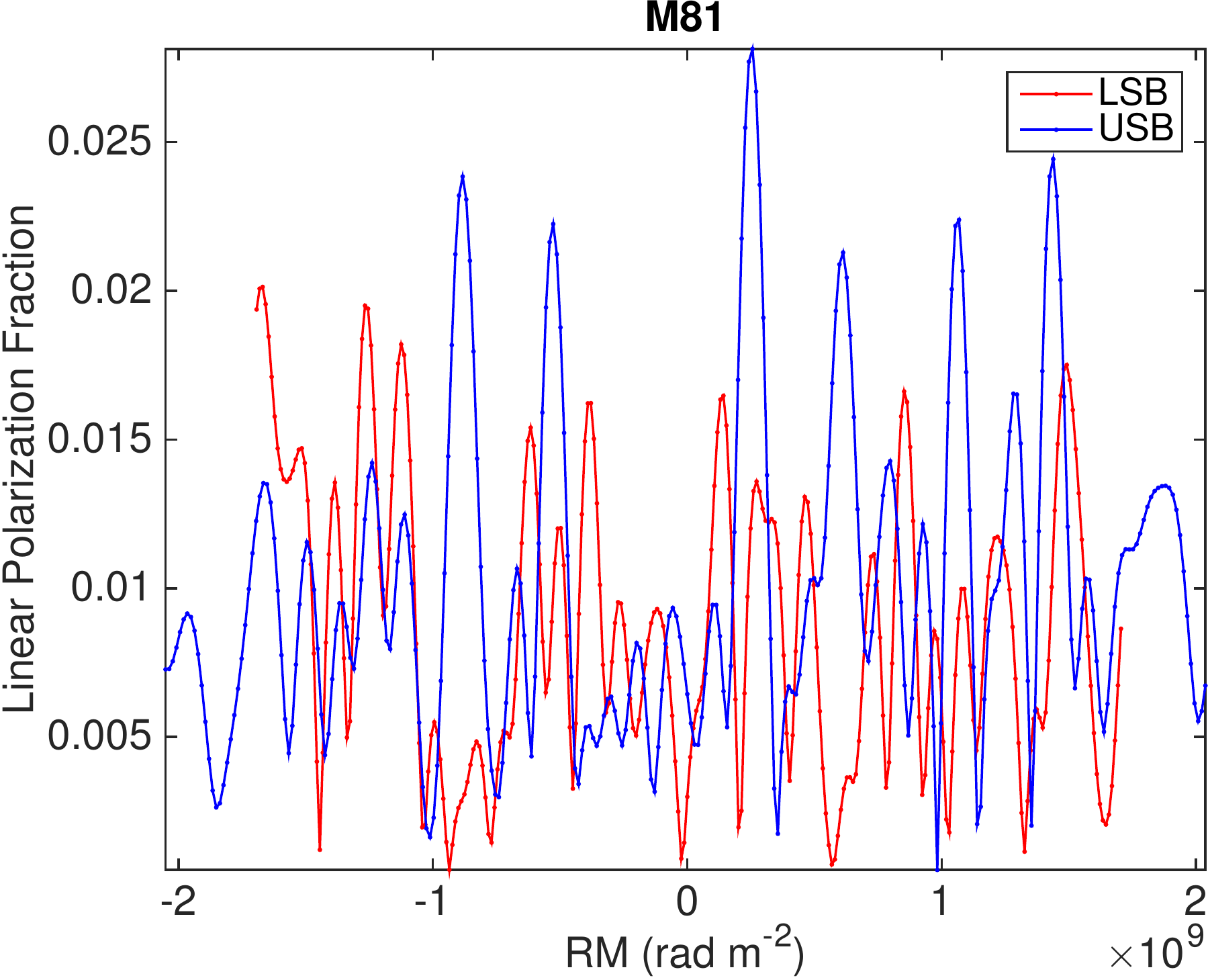}\includegraphics[width=0.5\textwidth]{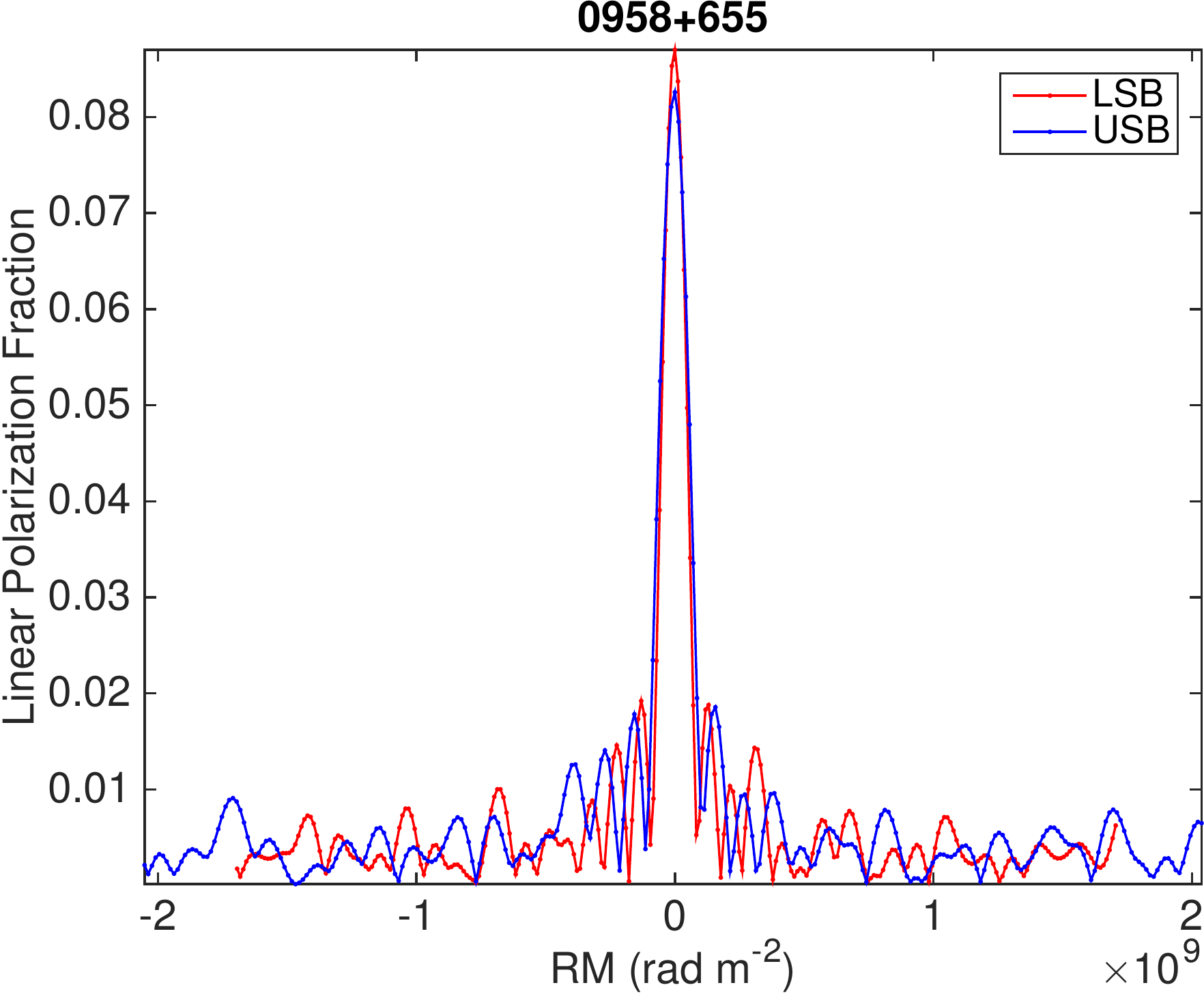}
\includegraphics[width=0.5\textwidth]{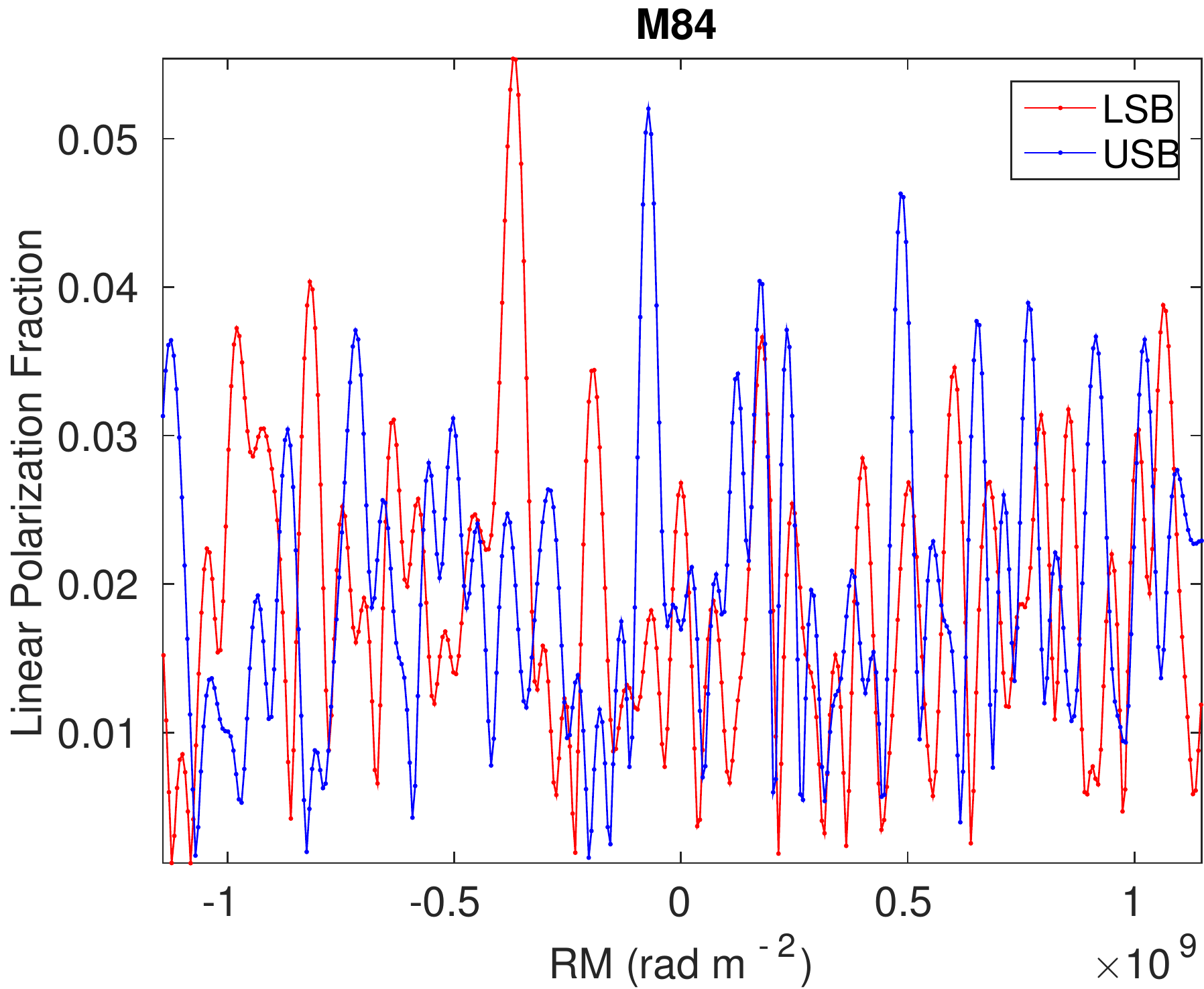}\includegraphics[width=0.5\textwidth]{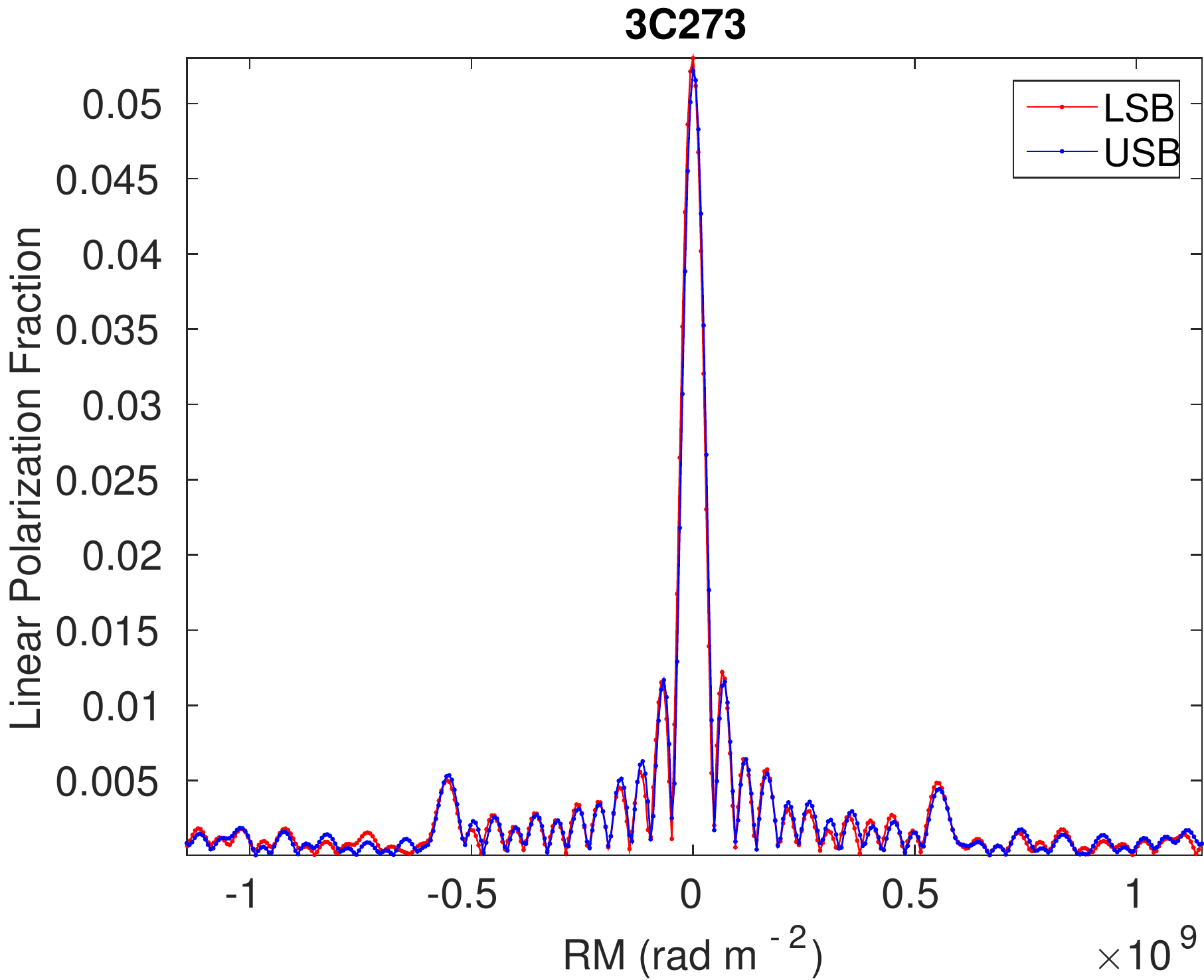}
\caption{Rotation measure synthesis results for M81 (upper left), M84 (lower left), 0958+655 (upper right), and 3C 273 (lower right).  Results show the linear polarization fraction as a function
of RM for the upper and lower sidebands.  The M84 results are from the most sensitive epoch, 160221.
The results for the calibrators 0958+655 and 3C 273
are consistent with ${\rm RM}=0 \rdm$.  No detection is made for either M81 or M84 in any epoch.
\label{fig:rmspectra}
}
\end{figure}

\section{Results and Analysis}

We summarize our polarimetric results for M81, M84, and their calibrators in Table~\ref{tab:results}.
Results are tabulated per epoch for each sideband in Stokes $I, Q,$ and $U$ parameters.
No polarized emission is detected from either M81 or M84.  There are no detections 
for either source in upper or lower sidebands or in the average of the two sidebands.
Further, M84 is not detected in any epoch, as well as not in the
average over all epochs and sidebands.  The calibrators 0958+655 and 3C273 are clearly detected
in all sidebands and epochs with consistent values.

We also searched spectral data for RMs up to a maximum
\begin{equation}
{\rm RM}_{lim} = {\pi \over 2 \lambda^2} {\nu \over \Delta\nu},
\end{equation}
where $\lambda$\ and $\nu$\ refer to observing wavelength and frequency, and $\Delta\nu$\ 
is the channel width \citep{2005A&A...441.1217B}.  The M81 CARMA data were spectrally averaged into 36 channels of width 0.052
GHz in each sideband.  The M84 SMA data were averaged in each of 48 spectral window of width 0.112
GHz. In the lower sideband, this spectral structure corresponds to ${\rm RM}_{lim}=3.4 \times 10^9 \rdm$ and $1.7 \times 10^9 \rdm$ for M81 and M84, respectively.  RM synthesis for the two calibrators revealed 
upper and lower sideband results both consistent with ${\rm RM}=0 \rdm$.  The width of the RM 
transfer function is $4.7 \times 10^7 \rdm$ and $2.3 \times 10^7 \rdm$ for M81 and M84, respectively. 
No significant peaks were 
found in the RM spectra for M81 and M84 in any epoch (Figure~\ref{fig:rmspectra}).  Upper limits are $\sim 2\%$ of the total intensity.

Under the assumption that M81 and M84 are intrinsically linearly polarized but appear
unpolarized due to bandwidth depolarization, 
we place two lower limits on the {\rm RM}.  The first is based on the non-detection in
either the lower or upper sideband averaged channels.  For the case of 1 radian of rotation, 
the RM limits are $3.1 \times 10^7 \rdm$
and $1.6 \times 10^7 \rdm$ for M81 and M84, respectively.  The second is based on
non-detection in the RM spectra with the limits given above, $\sim 10^9 \rdm$.
The fractional polarization limits from these two methods are comparable.

Whether the RM limit is physically applicable depends on the uniformity of the Faraday medium 
over the solid angle of the source.  In this case,
the large RM $\sim 10^9 \rdm$ implies $\sim 600$ complete turns of phase at our observational
wavelength.  Inhomogeneities of a part in $10^3$ in the Faraday medium could lead to beam depolarization.  
Models for Faraday rotation in the accretion flow discussed below are insufficiently detailed to 
provide any meaningful constraint on beam depolarization for the very compact size of the polarized emission.

\section{Discussion}

Millimeter wavelength radio-loud AGN have been shown to be strongly polarized
\citep{2010A&A...515A..40T,2014A&A...566A..59A}.  At 3.4 mm,
surveys have shown a median fractional polarization of $\approx 4$\%
with values ranging as high as 19\%. Polarized flux is detected in $\sim 90\%$ 
of sources at fractions above 1\%.   The average 3 mm polarization fraction is 
a factor $\sim 2$ times higher than the average polarization fraction at 2 cm.  
Surveys at 1.3 mm detect fewer 
polarized sources as a result of reduced sensitivity, but 
for those sources that are detected the mean polarization fraction increases
by $\sim 1.6$ times relative to 3.4 mm.  
The results are consistent with a general trend of increased magnetic
field order from the smaller source regions that are expected with
decreasing wavelength.

\begin{figure}[p!]
\includegraphics[width=\textwidth]{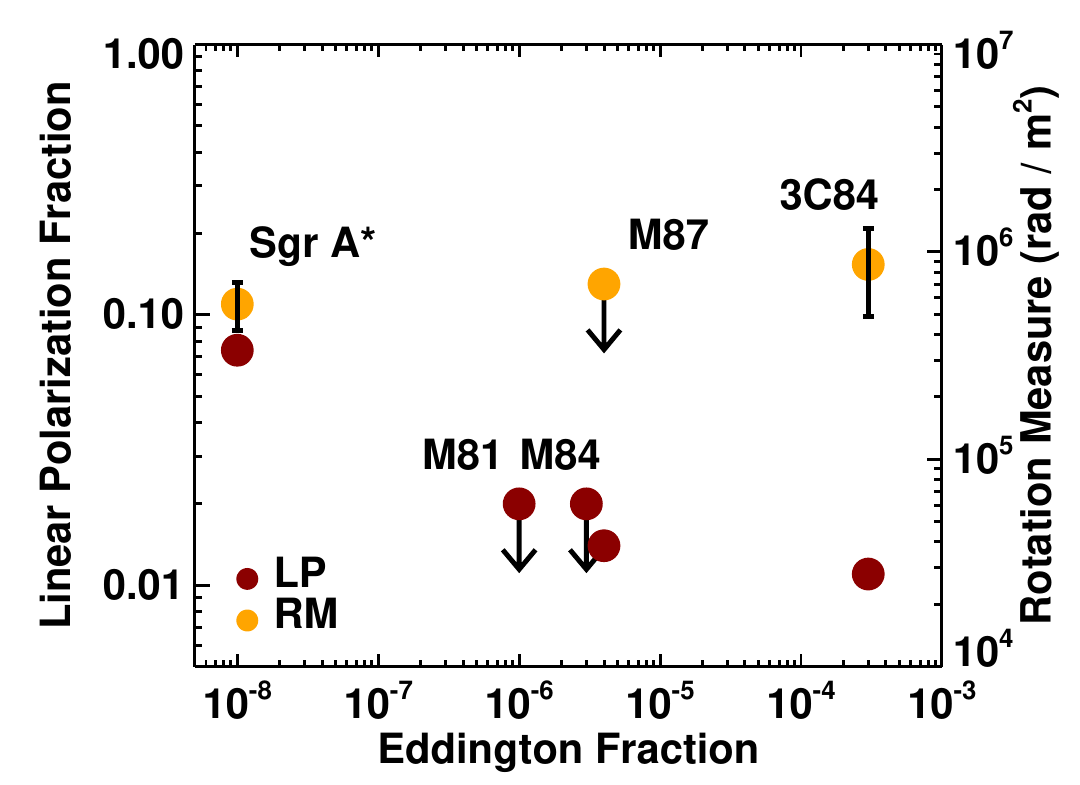}
\caption{Linear polarization fraction (LP, red circles) and rotation
  measure (RM, yellow circles) as a function of 
the Eddington fraction, i.e., 
the ratio of bolometric luminosity to Eddington luminosity.
The 3 available RM constraints show roughly constant values
  despite large differences in black hole mass and accretion rate. The
measured LPs except for Sgr A* are also low compared to radio loud
AGN. For M81 and M84, the low LP could be the result of beam or Faraday depolarization. 
\label{fig:polfrac}
}
\end{figure}

The non-detection of polarization from the LLAGN M81 and M84, as well as
the weak linear polarization detected from M87 and 3C 84, suggest that these sources
are qualitatively different from the blazars and radio
loud objects that make up the survey samples.  These differences may arise
from the very different radii at which the emission arises due to the beamed jet
structure or from the effect of jet orientation.  Further, the LLAGN
polarization fractions appear to be qualitatively different from that of 
Sgr A*, which shows $\sim 10\%$ polarization fraction at mm wavelengths
(Figure~\ref{fig:polfrac}).  This is in contrast with the consistent cm 
wavelength picture of the mm polarization of Sgr A* and M81:  both exhibit 
undetected linear polarization along with detectable circular polarization
\citep{2001ApJ...560L.123B}.  We consider below whether the mm polarization
differences can be accounted for with internal emission or extrinsic Farday effects, or
some combination thereof.

The large RM towards \sgra\ has been successfully modeled as the result of propagation through the
ionized, magnetized accretion flow \citep{2003ApJ...588..331B,2007ApJ...654L..57M}.  The RM is directly tied to the mass accretion
rate in these models.  Generically, models make predictions for the accretion rate using
radial profiles
in electron density, $n_e \propto r^{\beta}$, and magnetic field strength, which
is typically assumed to be  in equipartition with the particle kinetic energy.
ADAF models without winds or convection
have centrally peaked profiles ($\beta=3/2$) that have diminishing importance as
the source of the emission moves further out; RIAF models \citep[$\beta\approx 1/2$;][]{1999MNRAS.303L...1B} have a broad maximum in RM at larger radii and so are
less sensitive to the location of the emission region.  Exact RM predictions are
a function of black hole mass ($M_{bh}$), mass accretion rate at the inner
radius of the accretion flow ($\dot{M}$ in $M_\odot$ per year), and the inner
and outer radii of the accretion flow 
\citep[$r_{in}$, $r_{out}$ expressed in $R_s$;][]{2006JPhCS..54..354M}\footnote{Note the corrected sign on the exponent
for $r_{in}$.}:
\begin{equation}
{\rm RM} \propto \left( {2 \over 3\beta -1 } \right) \left( 1 - \left({r_{out} \over r_{in} }\right)^{-(3\beta-1)/2} \right)   r_{in}^{-7/4} M_{bh}^{-2} \dot{M}^{3/2}.
\label{eq:rm}
\end{equation}
These models work well for Sgr A*  and provide reasonable estimates for
the case of M87 \citep{2014ApJ...783L..33K} but underpredict the RM
observed towards 3C 84 by orders of magnitude \citep{2014ApJ...797...66P}.

If we assume that M81 and M84 are intrinsically polarized but depolarized through
the accretion medium, then we can place constraints on the accretion flow properties.
Both M81 and M84 are compact sources with jet structure marginally resolved 
at sub-parsec resolution
\citep{2000ApJ...532..895B,2004AJ....127..119L,2008ApJ...681..905M}.  In the case of
M84, 7mm VLBA observations show that the source is compact on a scale of few hundred $R_S$.
In the case of M81, a core with a compact jet is seen at a wavelength of 3.4 cm; the core has
a maximum scale of $10^3 R_S$.  If these sources resemble M87,
then we expect the polarization to arise from even more compact regions, comparable
to $\sim 10 R_S$.  Thus, it is reasonable to test the hypothesis that 
polarized emission is generated on small scales and then propagates
through a geometrically thick (quasi-spherical) accretion flow, where it is depolarized.

\begin{figure}[p!]
\includegraphics[width=0.5\textwidth]{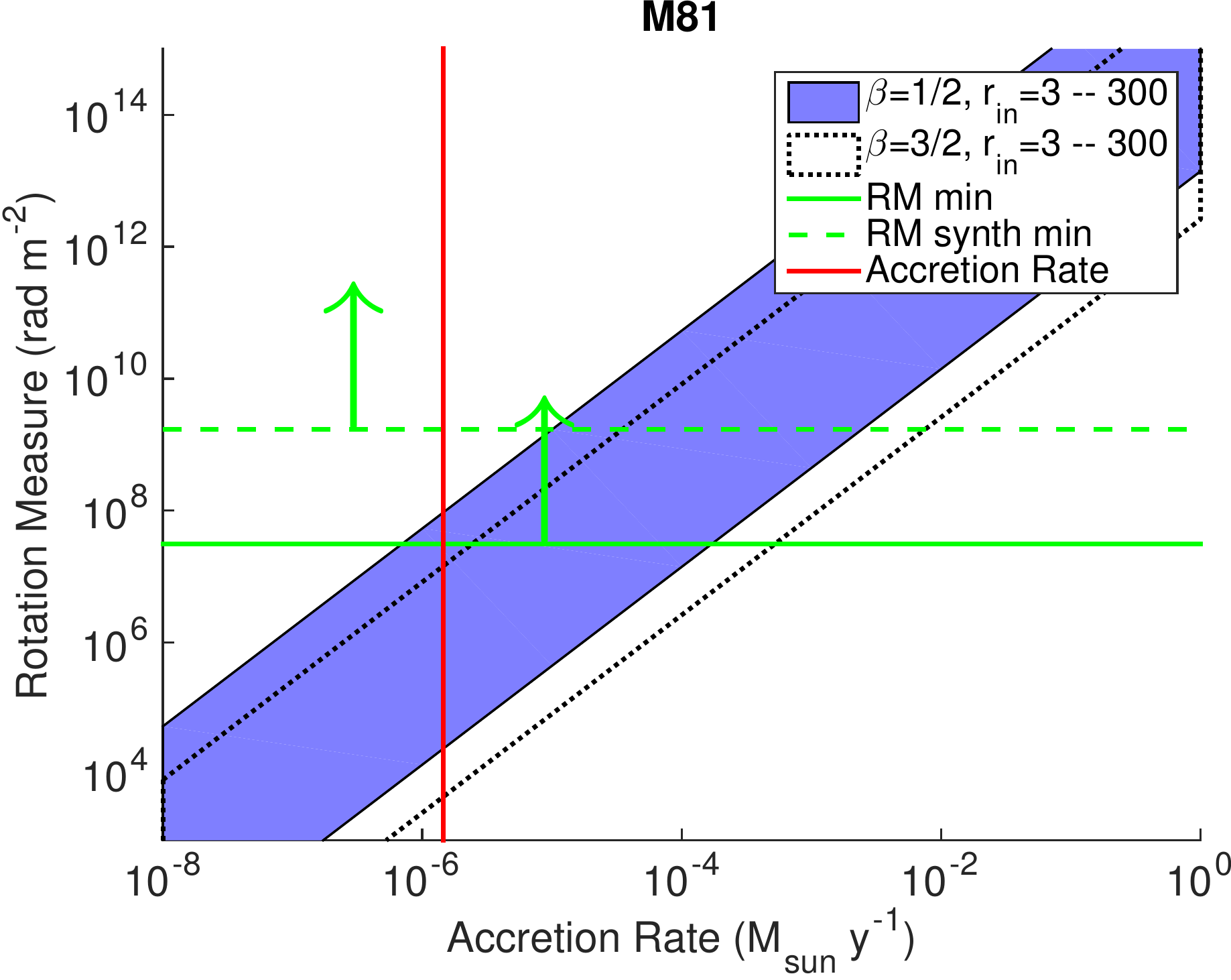}
\includegraphics[width=0.5\textwidth]{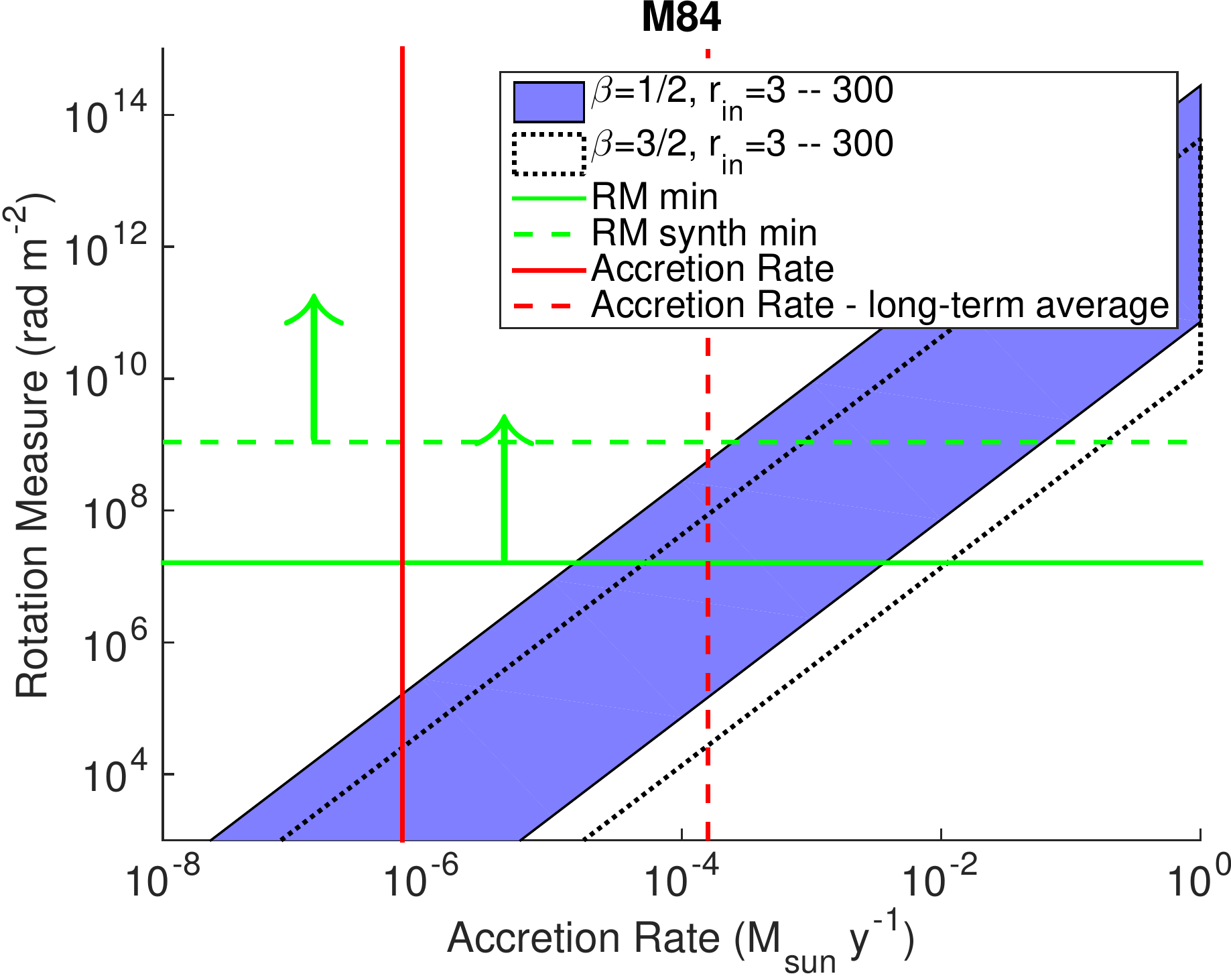}
\caption{Rotation measure versus accretion rate for M81 (left) and M84 (right).  The vertical
red line is the inferred accretion rate from the X-ray luminosity.
The horizontal green lines are the minimum RM to depolarize the source 
within each sideband (solid line) and within each channel (dashed line).
The blue shaded region represents parameter space associated with
RIAF models ($\beta=1/2$) for $r_{in} = 3$ to 300.  The space 
enclosed within the dotted lines represent parameter space associated with
ADAF models ($\beta=3/2$).  Space to the upper left of the model curves is
excluded.
Additionally, we show
a vertical red line indicating the long-term average for the M84
accretion rate.
\label{fig:rm}
}
\end{figure}

In Figures~\ref{fig:rm} we plot our two RM limits against mass accretion
rate, as well as expectations for ADAF and RIAF models with a range of inner and outer radii.
We obtain the mass accretion rate for M81 from detailed spectral modeling, which treats
the observed X-ray luminosity as primarily a non-thermal component associated with the jet
\citep{2008ApJ...681..905M}.  The various M81 models are consistent with an accretion disk
bolometric luminosity that is $10^{-6} L_{Edd}$, 
or an accretion rate of $10^{-6} \eta_{0.1}^{-1} M_\odot\ {\rm y^{-1}}$,
where $\eta_{0.1}$ is the efficiency normalized by the thin-disk
value of 0.1.  
We determine the instantaneous 
mass accretion rate for M84 based on the unabsorbed X-ray luminosity of the nucleus at $7 \times 10^{-7}\ M_\odot\ {\rm y^{-1}}$
\citep{2013MNRAS.432..530R}.  A long-term average accretion rate of 
$2 \times 10^{-4}M_\odot\ {\rm y^{-1}}$ 
for M84 has been also determined
through analysis of the power required to create the observed X-ray cavity, which
has a  time scale $\sim 10^6$ y \citep{2006ApJ...652..216R}.  
Both M81 and M84 are known to show variable nuclear X-ray
emission with up to an order of magnitude change in luminosity on time scales of months to years.
Thus, the accretion rate for both sources may be uncertain by more than an order of magnitude.

In the case of M81, we see depolarization in the accretion flow is marginally consistent
with RIAF and ADAF models.  Those models must have cold (non-relativistic) electrons down to small radii, $r_{in} 
\approx 3$, to produce sufficient depolarization.  Alternatively, clumpiness
at large radii or  a higher long-term 
accretion rate could lead these results to be consistent with accretion theory.
In the case of M84, for the current epoch accretion rate, both RIAF and ADAF models are 
strongly rejected.  For the long-term average accretion rate, however, RIAF and ADAF
models can be constructed that are consistent with non-detection of linear polarization.

It is unclear whether this simple picture of the intrinsically polarized source
and external Faraday medium  will hold in these sources.
First, Sgr A* reveals significant complexity in its polarization on $R_S$ scales,
possibly as the result of spatially-variable magnetic-field structure \citep{2015Sci...350.1242J}.
Second, long wavelength RM synthesis observations of Sgr A* that were sensitive to 
$RM \leq 1.5 \times 10^7 \rdm$
did not detect the polarization in spite of the fact that millimeter wavelength 
polarimetry detected a strongly polarized source at $RM = -5 \times 10^5 \rdm$ 
\citep{1999ApJ...521..582B}.  The absence of that detection is likely either due to 
beam depolarization for the much larger intrinsic source at long wavelengths, or
to other effects that suppress the intrinsic polarization.  In the case of Sgr A*, 
the polarized millimeter wavelength source may represent only the inner
accretion flow or jet structure, whereas the long wavelength emission may represent a 
larger-scale jet
or nonthermal particles in the larger-scale accretion flow.  On the other hand, for 
both M81 and M84, like M87,
the spectrum is consistent with a single source component that is associated with
a self-similar jet.

We also consider the possibility that the sources are not intrinsically polarized.  Given
the compact source size of the emitting regions at these wavelengths, estimated to be tens
of $R_S$, this requires a substantial degree of cancellation over a small area.  
Magnetic-field
structure in Sgr A* appears to lead to a reduction from the peak polarized fraction 
to the average polarized fraction by a factor of several.  Such a reduction still 
leaves a substantially polarized source.  Both M81 and M84 show a steep spectrum indicative
of optically thin emission, eliminating optically thick emission as a source for
internal depolarization in the mm.  A more plausible hypothesis is 
mixed thermal and nonthermal material within the emission region that leads to 
internal depolarization.

Numerical models using semi-analytic prescriptions for electron heating
\citep{2014A&A...570A...7M,2016A&A...586A..38M,2017MNRAS.467.3604R} can now
self-consistently calculate polarized emission from low-luminosity
accretion flows including internal and external Faraday effects at
least out to $r \sim 100 R_S$ \citep{2016MNRAS.462..115D,2017MNRAS.468.2214M}. The models can reproduce flat
radio spectra from a self-absorbed jet \citep[e.g.,][]{1979ApJ...232...34B,1995A&A...293..665F}. At 230 GHz the emission is predicted to
originate from close to the event horizon, where the counter-jet is
lensed into a crescent shape \citep{2012MNRAS.421.1517D}. The LP from this
configuration is typically only $\simeq 1-3\%$ as a result of either
beam or Faraday depolarization \citep{2017MNRAS.468.2214M}, consistent
with observations of M87 \citep{2014ApJ...783L..33K}.

In these numerical models,
mildly relativistic or cold electrons near the black hole produce strong Faraday
depolarization ($r_{\rm in} \sim R_S$) and large RM
values when viewed at moderate to high inclinations, roughly in
agreement with Equation~\ref{eq:rm}. When viewed at low inclination
(e.g. M87), the RM can be much lower than predicted, such that
$\gtrsim 10\times$ higher accretion rates are consistent with the observations. At
larger inclination, the RM is predicted to increase rapidly. The
non-detections of LP in M81 and M84 are consistent with the
\citet{2017MNRAS.468.2214M} model predictions if the 230 GHz emission
originates near the event horizon \citep[as suggested for M81 by its short variability timescale,][]{
2015ApJ...811L...6B}. Future higher sensitivity observations could
determine whether the RM is large (moderate to high inclination), or
similar to that measured for M87, possibly indicating horizon scale
counter-jet emission viewed at small inclination.

\section{Conclusions}

We have presented CARMA and SMA millimeter wavelength polarimetry measurements for the
LLAGN M81 and M84.  We detect no linear polarization from these sources in spite of the
sensitivity to very large RMs.  Interpreted through the model of a compact, 
intrinsically depolarized source that is depolarized by a homogeneous accretion flow that extends from close to the black hole to the Bondi radius, we find that
the results are inconsistent or marginally consistent with other constraints on the
accretion flow.  Additional cold, depolarizing material near the black hole may be
responsible for the depolarization.  

These results, along with those for Sgr A*, M87, and 3C 84, demonstrate that LLAGN have
different properties than we see for higher luminosity radio sources at these wavelengths.
Study of a broader sample of objects over a wide range of wavelengths and at
greater sensitivity with ALMA will be critical 
for understanding this tool for probing accretion flows and jets on scales of a few to 
thousands of Schwarzschild radii.  Finally, in the case of the brightest sources, polarized
imaging with the Event Horizon Telescope may produce maps of polarization structure
that cannot be obtained in any other way.

\begin{deluxetable}{lllrrrrrrrr}
\tablecaption{Polarimetric Results \label{tab:results}}
\tablehead{
\colhead{Tel.} & \colhead{Epoch} & \colhead{Source} & \colhead{$\nu$} & 
\colhead{$I$} &
\colhead{$Q$} &
\colhead{$U$} &
\colhead{$P$} \\
               & \colhead{(yymmdd)} &               & \colhead{(GHz)} &
\colhead{(mJy)} &
\colhead{(mJy)} &
\colhead{(mJy)} &
\colhead{(mJy)} 
}
\startdata
CARMA & 130218 &      M81 & 216.6 & $  75.0 \pm   0.6 $& $  -0.2 \pm   0.5 $& $   0.1 \pm   0.6 $& $   0.2 \pm   0.5 $ \\ 
\dots &  \dots &    \dots & 231.1 & $  77.3 \pm   0.8 $& $  -0.5 \pm   0.6 $& $   0.0 \pm   0.7 $& $   0.5 \pm   0.6 $ \\ 
\dots &  \dots &    \dots & Mean & $  76.1 \pm   0.5 $& $  -0.3 \pm   0.4 $& $   0.1 \pm   0.5 $& $   0.4 \pm   0.4 $ \\ 
CARMA & 130218 & 0958+655 & 216.6 & $ 698.8 \pm   1.0 $& $  60.8 \pm   1.1 $& $  -0.6 \pm   1.3 $& $  60.8 \pm   1.1 $ \\ 
\dots &  \dots &    \dots & 231.1 & $ 704.7 \pm   1.3 $& $  58.2 \pm   1.2 $& $   0.7 \pm   1.7 $& $  58.2 \pm   1.2 $ \\ 
\dots &  \dots &    \dots & Mean & $ 701.8 \pm   0.8 $& $  59.5 \pm   0.8 $& $   0.1 \pm   1.0 $& $  59.5 \pm   0.8 $ \\ 
\hline
SMA & 160130 &      M84 & 220.9 & $ 119.1 \pm   1.6 $& $  -1.7 \pm   1.3 $& $  -0.1 \pm   1.6 $& $   1.7 \pm   1.3 $ \\ 
\dots &  \dots &    \dots & 232.9 & $ 114.0 \pm   1.4 $& $  -2.3 \pm   1.4 $& $   0.5 \pm   1.3 $& $   2.3 \pm   1.4 $ \\ 
\dots &  \dots &    \dots & Mean & $ 116.5 \pm   1.1 $& $  -2.0 \pm   1.0 $& $   0.2 \pm   1.0 $& $   2.0 \pm   0.9 $ \\ 
SMA & 160130 &    3C273 & 220.9 & $ 9600.2 \pm   4.3 $& $  67.1 \pm   2.6 $& $ -382.0 \pm   3.6 $& $ 387.8 \pm   3.5 $ \\ 
\dots &  \dots &    \dots & 232.9 & $ 9872.9 \pm   3.1 $& $ 102.1 \pm   2.8 $& $ -317.3 \pm   5.7 $& $ 333.3 \pm   5.5 $ \\ 
\dots &  \dots &    \dots & Mean & $ 9736.5 \pm   2.5 $& $  84.6 \pm   1.9 $& $ -349.7 \pm   3.0 $& $ 359.7 \pm   3.0 $ \\ 
SMA & 160221 &      M84 & 220.9 & $ 129.4 \pm   1.9 $& $   2.1 \pm   1.9 $& $  -2.7 \pm   2.1 $& $   3.5 \pm   2.0 $ \\ 
\dots &  \dots &    \dots & 232.9 & $ 144.9 \pm   1.8 $& $  -2.4 \pm   2.5 $& $  -0.3 \pm   2.1 $& $   2.5 \pm   2.5 $ \\ 
\dots &  \dots &    \dots & Mean & $ 137.2 \pm   1.3 $& $  -0.1 \pm   1.5 $& $  -1.5 \pm   1.5 $& $   1.5 \pm   1.5 $ \\ 
SMA & 160221 &    3C273 & 220.9 & $ 12298.1 \pm  49.3 $& $ 339.5 \pm   4.2 $& $ -556.5 \pm   6.3 $& $ 651.9 \pm   5.8 $ \\ 
\dots &  \dots &    \dots & 232.9 & $ 11179.6 \pm  19.9 $& $ 317.2 \pm   4.3 $& $ -489.4 \pm   5.7 $& $ 583.2 \pm   5.4 $ \\ 
\dots &  \dots &    \dots & Mean & $ 11738.8 \pm  18.5 $& $ 328.4 \pm   3.0 $& $ -523.0 \pm   4.2 $& $ 617.5 \pm   3.9 $ \\ 
SMA & 160229 &      M84 & 220.9 & $ 160.0 \pm   1.5 $& $  -0.9 \pm   1.3 $& $  -2.0 \pm   1.7 $& $   2.2 \pm   1.7 $ \\ 
\dots &  \dots &    \dots & 232.9 & $ 171.3 \pm   1.6 $& $  -2.7 \pm   1.6 $& $  -1.6 \pm   1.6 $& $   3.1 \pm   1.6 $ \\ 
\dots &  \dots &    \dots & Mean & $ 165.7 \pm   1.1 $& $  -1.8 \pm   1.0 $& $  -1.8 \pm   1.2 $& $   2.5 \pm   1.1 $ \\ 
SMA & 160229 &    3C273 & 220.9 & $ 10936.5 \pm   3.4 $& $ 368.1 \pm   2.8 $& $ -556.8 \pm   3.8 $& $ 667.5 \pm   3.5 $ \\ 
\dots &  \dots &    \dots & 232.9 & $ 11202.3 \pm   3.9 $& $ 380.1 \pm   4.0 $& $ -577.6 \pm   4.0 $& $ 691.5 \pm   4.0 $ \\ 
\dots &  \dots &    \dots & Mean & $ 11069.4 \pm   2.6 $& $ 374.1 \pm   2.3 $& $ -567.2 \pm   2.8 $& $ 679.5 \pm   2.6 $ \\ 
\hline
SMA & Mean & M84 & Mean & $ 139.3 \pm  22.9 $ & $  -1.4 \pm   0.6 $ & $  -0.8 \pm   0.7 $ & $   1.6 \pm   0.6 $ \\ 
\enddata
\end{deluxetable}


\end{document}